\documentclass[conference]{IEEEtran}
\IEEEoverridecommandlockouts

\usepackage{cite}
\usepackage{amsmath,amssymb,amsfonts}
\usepackage{algorithmic}
\usepackage{graphicx}
\usepackage{textcomp}
\usepackage{xcolor}
\usepackage{url}
\usepackage{booktabs}
\usepackage{tikz}

\def\BibTeX{{\rm B\kern-.05em{\sc i\kern-.025em b}\kern-.08em
    T\kern-.1667em\lower.7ex\hbox{E}\kern-.125emX}}

\newcommand\copyrighttext{%
  \scriptsize \textcopyright~2025 IEEE. Personal use of this material is permitted.  Permission from IEEE must be obtained for all other uses, in any current or future media, including reprinting/republishing this material for advertising or promotional purposes, creating new collective works, for resale or redistribution to servers or lists, or reuse of any copyrighted component of this work in other works. doi: {10.1109/AIxVR63409.2025.00037}}
\newcommand\copyrightnotice{%
\begin{tikzpicture}[remember picture,overlay]
\node[anchor=south,yshift=10pt] at (current page.south) {\fbox{\parbox{\dimexpr\textwidth-\fboxsep-\fboxrule\relax}{\copyrighttext}}};
\end{tikzpicture}%
}

\begin{document}

\title{CUIfy the XR: An Open-Source Package to Embed LLM-powered Conversational Agents in XR \\
\thanks{We thank TUM Think Tank for supporting this work.}
}

\author{
\author{\IEEEauthorblockN{Kadir Burak Buldu$^1$\IEEEauthorrefmark{1}, Süleyman Özdel$^1$\IEEEauthorrefmark{2}, Ka Hei Carrie Lau$^1$\IEEEauthorrefmark{3}, Mengdi Wang$^1$\IEEEauthorrefmark{4}, Daniel Saad$^1$\IEEEauthorrefmark{5}, Sofie Schönborn$^1$\IEEEauthorrefmark{6}, \\ Auxane Boch$^1$\IEEEauthorrefmark{7}, Enkelejda Kasneci$^1$\IEEEauthorrefmark{8}, and Efe Bozkir$^1$\IEEEauthorrefmark{9}}
\IEEEauthorblockA{$^1$ Technical University of Munich (TUM), Germany \\
\IEEEauthorrefmark{1} burak.buldu@tum.de
\IEEEauthorrefmark{2} ozdelsuleyman@tum.de
\IEEEauthorrefmark{3} carrie.lau@tum.de
\IEEEauthorrefmark{4} mengdi.wang@tum.de
\IEEEauthorrefmark{5} daniel-saad@tum.de \\
\IEEEauthorrefmark{6} sofie.schoenborn@tum.de
\IEEEauthorrefmark{7} auxane.boch@tum.de
\IEEEauthorrefmark{8} enkelejda.kasneci@tum.de
\IEEEauthorrefmark{9} efe.bozkir@tum.de
}
}
}

\maketitle
\copyrightnotice

\begin{abstract}
Recent developments in computer graphics, machine learning, and sensor technologies enable numerous opportunities for extended reality (XR) setups for everyday life, from skills training to entertainment. With large corporations offering affordable consumer-grade head-mounted displays (HMDs), XR will likely become pervasive, and HMDs will develop as personal devices like smartphones and tablets. However, having intelligent spaces and naturalistic interactions in XR is as important as technological advances so that users grow their engagement in virtual and augmented spaces. To this end, large language model (LLM)--powered non-player characters (NPCs) with speech-to-text (STT) and text-to-speech (TTS) models bring significant advantages over conventional or pre-scripted NPCs for facilitating more natural conversational user interfaces (CUIs) in XR. This paper provides the community with an open-source, customizable, extendable, and privacy-aware Unity package, CUIfy, that facilitates speech-based NPC-user interaction with widely used LLMs, STT, and TTS models. Our package also supports multiple LLM-powered NPCs per environment and minimizes latency between different computational models through streaming to achieve usable interactions between users and NPCs. We publish our source code in the following repository: \textit{\url{https://gitlab.lrz.de/hctl/cuify}}
\end{abstract}


\section{Introduction}
Recent advances in computer graphics, hardware, and artificial intelligence may lead virtual and augmented reality (VR/AR) systems to become ubiquitous and head-mounted displays (HMDs) to be used more regularly. VR and AR have different use contexts and configurations, each providing distinct advantages for users. For instance, VR is beneficial for generating simulations to train users in a fully immersive setting~\cite{SAP19_safety_critical,Checa2020,Makransky_and_Klingenberg_2022,ooi_IEEE_VR_2021}; AR is more convenient when context-aware visual support is overlaid on real-world content~\cite{bektas_etal_2024,Ajanki2011,kim_etal_2018}. While both VR and AR, which are part of the broader field of extended reality (XR), can be useful for everyday life, making such virtual and augmented spaces intelligent often requires significant engineering effort, especially for environmental and social interactions. 

Considering intelligent XR spaces, practitioners often utilize non-player characters (NPCs), and these characters interact with users for different purposes~\cite{dobre_etal_2022,Zargham_etal_2024}. However, single-purpose NPCs may cause users to lose interest after a few interactions, as these characters tend to repeat the same or very similar content, eventually leading users to stop using the XR application. To this end, generative artificial intelligence (AI) and particularly large language models (LLMs) can provide numerous opportunities for XR due to their versatile computational capabilities, as they are trained with a significant portion of the Internet and can generate highly realistic synthetic data. By default, LLMs are utilized for the next-word prediction task; however, they can also be aligned for conversational purposes~\cite{touvron2023llama2openfoundation}, and one of the most prevalent examples is ChatGPT, which became publicly available in 2022~\cite{chatgpt_public_2022}. Since then, the general public has been more heavily exposed to generative AI systems and models. At the same time, the use of LLMs has accelerated in various domains, including medicine~\cite{li2023chatdoctor}, law~\cite{Cheong_etal_2024}, and education~\cite{hou_etal_2024}. 

While LLMs and generative AI models can be embedded into XR for different purposes, such as for creating 3D virtual content based on user preferences or for modifying interactive experiences, one of the most straightforward schemes is to embed LLMs into NPCs for speech-based interaction so that they can work as conversational user interfaces (CUIs). Prior research has utilized LLMs for speech-based NPC-user interactions with speech-to-text (STT) and text-to-speech (TTS) models~\cite{shoa_etal_2023,lau2024evaluating,Hajahmadi_etal_2024}. However, to the best of our knowledge, no open-source software implements the pipeline consisting of STT, LLM, and TTS models for XR in a generic and extendable way. The lack of such software means that for every XR application that includes LLM-based speech interaction, especially with NPCs, practitioners either implement the aforementioned pipeline from scratch or replicate it from previous projects, likely by also carrying out some modifications. 

Addressing the issue above, in this paper, we provide the community with an open-source Unity\footnote{Unity is a widely used, cross-platform game engine that Unity Software Inc. has developed.} package that combines LLMs, STT, and TTS models into a pipeline to enable speech-based interaction. Our package minimizes the latency between the models by utilizing streaming, supports plugging different models and pipelines into multiple NPCs in a single environment, and can prompt the LLMs. Furthermore, it supports accessing LLMs via application programming interfaces (APIs) and handling open-source LLMs on local devices or a separate server. Any Unity application can utilize our package; however, XR environments are particularly well-suited for its capabilities, enabling immersive and interactive experiences. We provide the source code in the following repository: \textit{\url{https://gitlab.lrz.de/hctl/cuify}}.

\section{Related Work} 
We divide this section into three subsections to cover the main aspects of interaction techniques in XR, LLM integration into XR, and existing LLM-based open-source tools that support user interaction in XR.

\subsection{Interaction in XR}
Interaction techniques and modalities for desktop interfaces (e.g., mouse and physical keyboard) are unsuitable for the immersive virtual interactions in XR~\cite{rauschnabel2022xr}. Various techniques with controllers, hand gestures, gaze, speech, or their combinations exist to interact in 3D immersive spaces~\cite{hou2021comparison, saktheeswaran2020touch}. Each method has distinct advantages, with speech-based interaction often integrated into other techniques in multimodal systems to enhance immersion and create intuitive XR experiences~\cite{wang2021interaction}.

Controller-based interaction is among the most common techniques in virtual spaces, offering easy-to-adapt input devices~\cite{caggianese2019vive}. Similarly, hand gesture-based interactions utilizing camera-based tracking systems~\cite{naguri2017recognition,ikram2020skeleton} or wearable technologies such as gloves~\cite{yang2019gesture} have gained importance~\cite{gavgiotaki2023gesture}. These methods enable natural, intuitive hand movements and enhance immersion by mimicking real-world interactions. Additionally, gaze-based interactions have been extensively studied as hands-free mechanisms for object selection and control~\cite{tanriverdi2000interacting,jalaliniya2014head,hansen2018fitts,blattgerste2018advantages,hulsmann2011comparing}.

Speech-based interaction is another method that provides a hands-free alternative, allowing users to interact naturally with the environment~\cite{zargham2024let,zargham2020smells}, especially in scenarios where a physical input is impractical or when users are engaged with other tasks~\cite{spittle2022review}. Those interactions offer higher adoption rates and better usability, particularly for novice users. Recent developments in LLMs can transform speech-based interaction in XR, as they provide versatile conversational capabilities. 

\subsection{Large Language Models in XR}
Interaction techniques in XR have rapidly evolved with different sensing modalities. Recently, with the spread of LLMs and their applications in various domains, LLM-powered NPCs and interactive objects in XR spaces have started to enable immersive and intuitive experiences. These advances enhance user interaction through verbal inputs, offering a hands-free, conversational means of engaging with the presented content. To this end, Bozkir et al.~\cite{bozkir2024embedding} argued for integrating LLMs into XR, emphasizing their potential for enhancing inclusion and engagement while raising concerns about the privacy of voice-enabled interactions. In another work, Liu et al.~\cite{liu2024classmeta} presented ClassMeta, LLM-driven interactive virtual classmates, in which the system uses voice commands to encourage student participation in virtual classrooms. The authors demonstrated the capabilities of LLMs in creating dynamic and interactive learning environments that simulate peer interactions. Additionally, Izquierdo-Domenech et al.~\cite{izquierdo2024virtual} combined VR with voice-enabled LLMs to provide context-aware educational experiences, significantly improving learning outcomes through personalized and natural interactions. 

Lau et al.~\cite{lau2024wrapped, lau2024evaluating} utilized VR and generative AI to enhance cultural heritage education by combining narrative personalization with LLM-powered chatbots. The authors demonstrated the potential of such configurations for improving user engagement, interactivity, and learning outcomes. Additionally, LLM-powered chatbots offered more dynamic learning experiences with higher usability than pre-scripted chatbots. Beyond speech-based interactions with LLMs, De et al.~\cite{de2024llmr} introduced the LLM for Mixed Reality (MR) framework, enabling real-time creation and modification of MR experiences using LLMs. Incorporating scene understanding, task planning, and self-debugging, it achieved a lower error rate than standard GPT-4 and received positive usability feedback, demonstrating its effectiveness.

\subsection{LLM-based Interaction Tools for XR}
Several open-source tools for XR have emerged to facilitate intuitive interactions. To this end, the OpenAI-Text-To-Speech-for-Unity~\cite{newtool1} package provides text-to-speech functionality in Unity using OpenAI's models. Similarly, Voice2Action~\cite{voice2action,su2023voice2action} enables voice-driven object manipulation in Unity. It integrates LLMs to allow users to adjust 3D objects with voice commands, making it possible to resize and reposition buildings in virtual environments. Another tool, LLMUnity~\cite{llmunity}, supports the implementation of LLM-powered characters in Unity, enhancing real-time interaction by enabling dynamic conversational agents that respond to user input, supporting local LLMs. Furthermore, Talk-With-LLM-In-Unity~\cite{talkwithllm} combines speech recognition with LLMUnity~\cite{llmunity} and the Google Gemma 2 model~\cite{team2024gemma}, enabling natural language-driven navigation. This package was primarily designed to integrate voice-controlled navigation and interaction in virtual environments; however, it does not support speech-based feedback. 

The EdenAI Unity Plugin~\cite{edenai} offers access to various AI services from Unity projects, including TTS, translation, and sentiment analysis. However, it is a commercial tool and does not provide a complete framework for NPCs. Additionally, it raises concerns over costs and privacy as it is a third-party service. While the aforementioned tools provide partial solutions for LLM-based interactions, none of them offer a comprehensive framework for NPCs that includes LLMs, STT, and TTS models. The previous solutions also do not allow selecting models separately for each service, which is essential for flexibility and privacy-sensitive use cases. Considering these, we built a generic and extendable CUIfy package for Unity.

\begin{table*}[t]
\caption{Overview of supported models and APIs. The Streaming column indicates whether streaming is supported, and the Local column specifies whether the model is local.}
\centering
\small
\begin{tabular}{c c c c c c}
\toprule
\textbf{Name} & \textbf{Speech to Text} & \textbf{Large Language Model} & \textbf{Text to Speech} & \textbf{Streaming} & \textbf{Local} \\ 
\midrule
OpenAI & \begin{tabular}[c]{@{}c@{}}Whisper\\ Whisper-tiny (local)\end{tabular} & \begin{tabular}[c]{@{}c@{}}GPT 3.5\\ GPT 4\\ GPT 4o\\ GPT 4o-mini\end{tabular} & TTS & \checkmark & $\times$ \\ \midrule
Amazon & Transcribe & $\times$ & Polly & \checkmark & $\times$ \\ \midrule
Google & $\times$ & \begin{tabular}[c]{@{}c@{}}Gemini 1.0 Pro \\ Gemini 1.5 Pro \\ Gemini 1.5 Flash\end{tabular} & $\times$ & \checkmark & $\times$ \\ \midrule
Meta & MMS-ASR & LLaMa (local) & MMS-TTS & \checkmark & \checkmark \\ \midrule
Hugging Face & \checkmark & \checkmark & \checkmark & \checkmark & \checkmark 
\\ \bottomrule
\end{tabular}
\label{tab:supported_APIs}
\end{table*}

\section{System Description}
We propose CUIfy, an open-source package that combines a backend server with Unity clients to provide easy-to-use LLM-powered conversational agents in Unity. Our package offers a user-friendly interface for different speech-to-text, text-to-speech, and large language models. It also provides options to select TTS voice types and allows users to utilize system prompts for LLMs. We discuss the technical specifications and system usage in the following subsections. 

\subsection{Technical Specifications}
The CUIfy package consists of a Python server to process voice input and a Unity client to create requests. The Unity client includes a user-friendly configuration interface. Users can choose various local or online models (i.e., through APIs) with different configuration options, listed in Table~\ref{tab:supported_APIs}. 

The Python server creates separate threads for each incoming socket connection and handles connections simultaneously. Each connection starts with incoming configuration messages, and the server can serve different APIs with different settings for each connection. These simultaneous connections enable different NPCs with various configurations to work in Unity environments with a single server. Each NPC can have a unique voice, system prompt, and conversation history.

The server uses state-of-the-art STT, TTS, and LLM APIs and models, all integrated via their official Python repositories to ensure compatibility and maintainability. Additionally, we used the latest versions of these APIs and models during testing and in our package to ensure up-to-date performance and reproducibility as depicted in Table~\ref{tab:supported_APIs}. The server is modular, making it straightforward to integrate both individual or publicly available models, whether local or online. Using Docker containers ensures that the package provides cross-platform operation without compatibility issues. 

The Unity client uses built-in Unity microphone API and .NET Socket Class~\cite{microsoft_socket} to ensure functionality on every native platform. Each object with the client script creates a unique socket connection with the server, allowing for the creation of multiple configurable NPCs. Furthermore, the server streams the LLM outputs to the client sentence-by-sentence if the model (and API) is supported and selected by the user. The server is configurable to run in the cloud environment with the correct network setup. The Dockerization process ensures service on any cloud server without struggling with dependencies. The client works seamlessly on Android, Windows, and macOS platforms, with the host on the same or different machine. We tested the CUIfy with Varjo XR-3, Meta Quest 2, and Meta Quest 3 HMDs using Unity version 2022.3.26f1, which is a Long Term Support version.

\begin{figure}[ht!]
    \centering
    \includegraphics[width=1.0\linewidth]{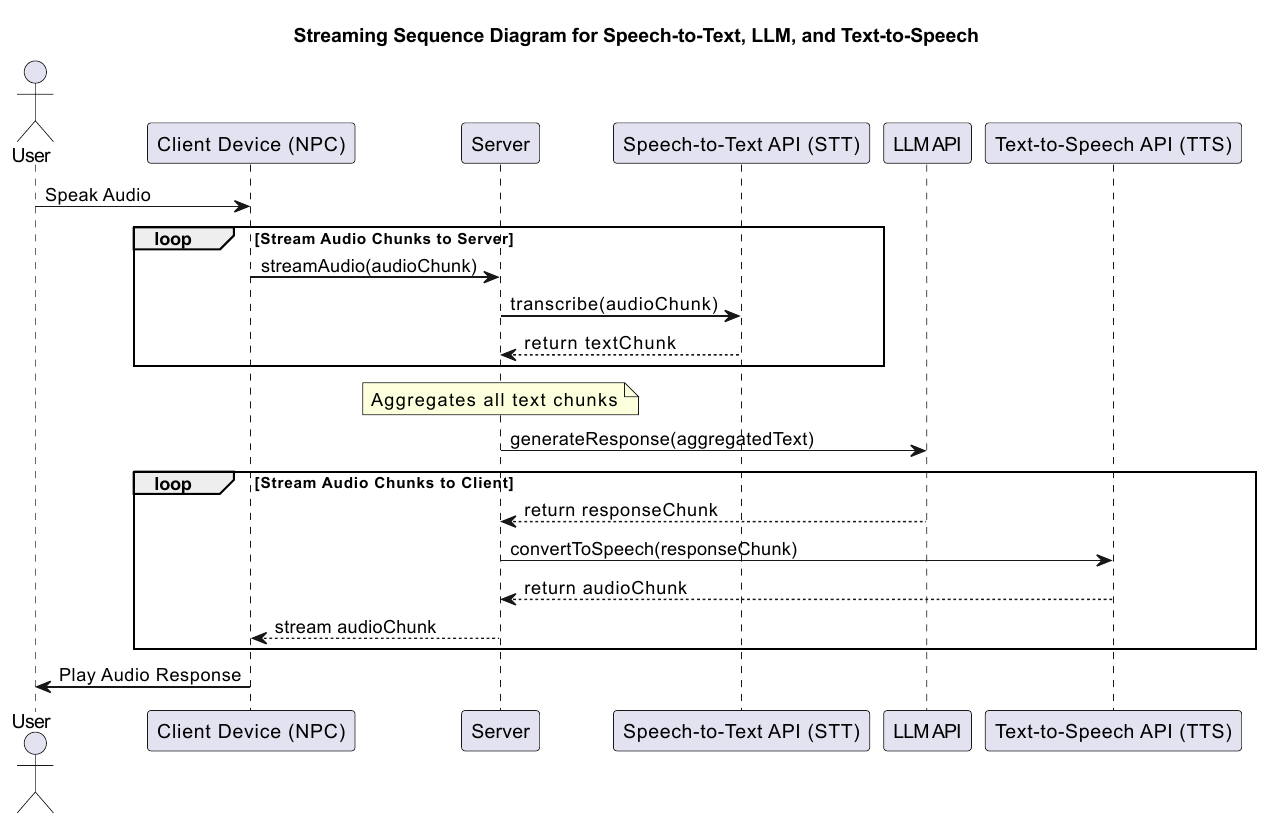}
    \caption{Sequence diagram of streaming pipeline.}
    \label{fig:CUIfy_sequence}
\end{figure}

Figure~\ref{fig:CUIfy_sequence} illustrates the streaming process used by CUIfy for STT, LLM, and TTS interactions in streaming mode. Audio input from the user is streamed to the server in chunks, where it is processed using the selected STT model to produce text chunks. These are aggregated and sent to the LLM for response generation. The LLM's output, returned in chunks, is processed by the TTS model into audio chunks, which are streamed back to the Unity client in real time for playback.

\subsection{System Usage and Guidelines} 
CUIfy's server supports and utilizes Docker containers, allowing an easy setup process. The source code is also well-designed and flexible, allowing users to add or edit models easily. The Unity package consists of a client script and additional libraries for working with audio data. It offers an interface in the Unity inspector section. It is possible to configure all the APIs, API keys, and other settings through the inspector. Users can select different models for processing and voices for different NPCs in the same Unity environment. 

Users should import the Unity package and run the Python server in Docker or their working environment to get started. The source code and the assets are available in the provided repository. Once the package is added to Unity, the client script can be attached to any object. It then needs to be adjusted to how the event will be triggered, such as assigning a button or colliding with an object. 

\begin{figure}[ht!]
    \centering    
    \includegraphics[width=0.599\linewidth]{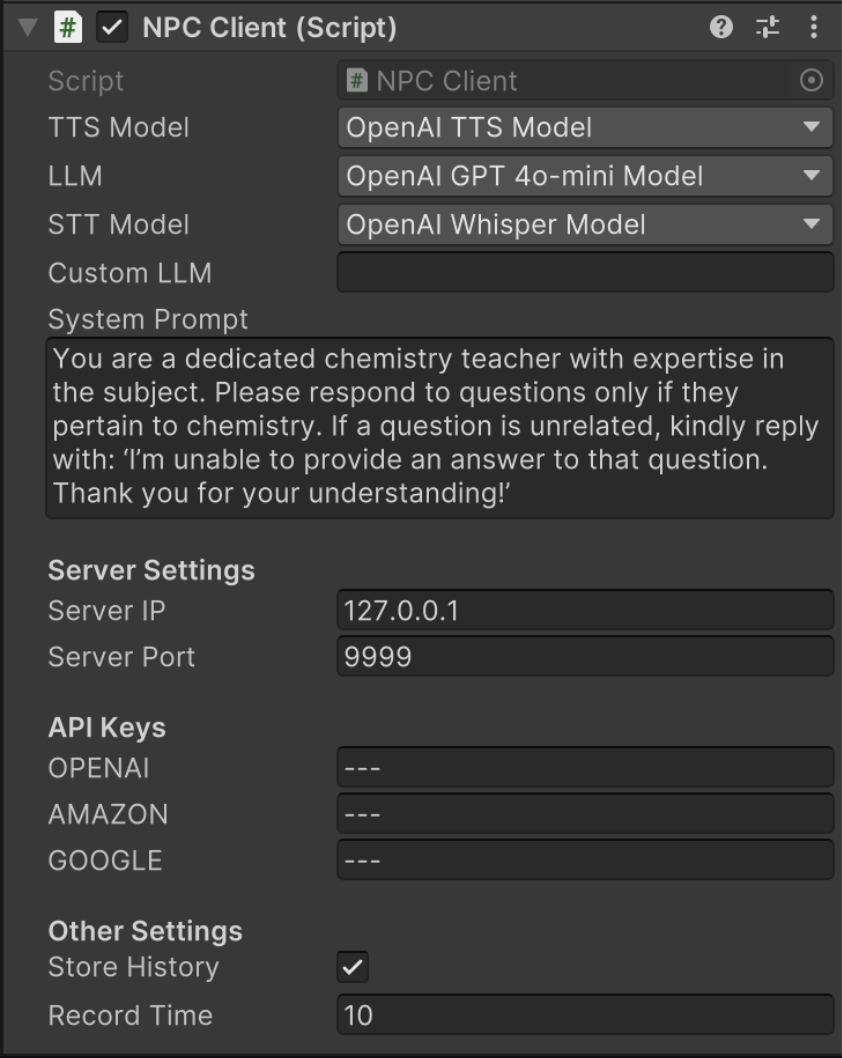}
    \caption{A sample view from the Unity inspector.}
    \label{fig:unity_inspector}
\end{figure}

CUIfy supports various online and local models by default. Since the backend server is modular, adding a new model or API object will suffice to extend the number of supported models. CUIfy currently supports mainstream APIs and local models, and Table~\ref{tab:supported_APIs} lists these. Users can select the desired model, enable the streaming mode, use their own API keys, and provide system prompts for LLMs from the Unity editor as depicted in Figure~\ref{fig:unity_inspector}. In addition, the Hugging Face Transformers pipeline~\cite{wolf2020transformers} is supported, and users can run any LLM model supported by Hugging Face on their local or cloud environment. By default, the server stores communication history between the server and clients. However, users can disable chat history for specific NPCs by unchecking its option in Unity Inspector, making each interaction feel new. 

Real-time communication with NPCs is essential for immersive experiments in XR as it builds a social presence for the user, the feeling that they are interacting with another person~\cite{LEGE2024100062}. CUIfy's streaming mode enables TTS and STT streaming instead of interrupting the computational process until all the audio or text data is available. In addition, when supported by the LLM API, instead of waiting for the entire LLM output, the TTS API will generate the audio data and send it to the client sentence by sentence. This streaming process ensures that the response time for the whole process is handled in real time with low latency. 

Investigating the outputs of LLMs is critical because these models, despite their powerful capabilities, can generate misinformation or biased content. Thus, it is essential to scrutinize their outputs to ensure accuracy, fairness, and reliability. Additionally, analyzing their behavior helps improve the models' alignment with ethical standards and user expectations~\cite{bender2021}. CUIfy provides detailed logs for each conversation conducted with LLMs. Users can, therefore, investigate and tune their system prompts to ensure NPCs respond as intended.

\section{Discussion and Future Directions}
The advancements in LLMs and speech models hold substantial potential for developing speech-based interaction techniques for virtual environments. They can also facilitate personalized, interactive NPCs and dynamic narrations within virtual spaces. In addition, these settings may enhance collaboration and accessibility and provide features such as real-time translations and transcriptions. With their inherently immersive nature, XR settings are especially well-suited for verbal interactions with NPCs. CUIfy leverages these potentials by providing a tool that simplifies the integration of speech-based NPCs into Unity environments, enabling natural and realistic interactions with minimal technical effort. This feature also benefits users without coding expertise, allowing an easy-to-use implementation in new and existing projects.

Key challenges like latency and conversation quality often arise in real-time conversational interactions. To address the latency issue, CUIfy employs a streaming technique that allows NPCs to speak as soon as content generation starts without waiting for the entire output. This approach significantly reduces the latency and may significantly improve user experience. While APIs and large local models often offer the best conversation quality, using APIs follows a pay-as-you-go model, and large models require considerable processing power. Our package also supports lightweight models that can run on local devices, though they generally provide slightly lower quality than full-scale models. Currently, CUIfy supports eight LLMs, four STT, and three TTS models, as provided in Table~\ref{tab:supported_APIs}. As generative AI and LLMs continue to evolve and new models emerge, CUIfy’s architecture allows the integration of those models and APIs easily, which remains the focus of future work. 

\section{Conclusion}
We introduced a generic, easy-to-use, and extendable Unity package that enables LLM-powered interactive NPCs designed for XR. Our package supports several LLMs, STT, and TTS models, allowing users to select models based on their preferences and needs while addressing quality and privacy requirements. Our package is optimized to enhance conversation quality through streaming and facilitate project integration by supporting APIs and local models. In future work, we plan to expand the number of supported models and conduct additional performance optimizations. 

\bibliographystyle{IEEEtran}
\bibliography{IEEEabrv,refs}

\end{document}